\documentclass[twoside,reqno]{HERON}
\usepackage{epsfig,cite,colordvi}
\usepackage{graphicx}
\usepackage{url}
\usepackage{amssymb,amsmath,amscd,epsf}
\usepackage{times}
\usepackage{makeidx}
\begin{document}

\title{Physics Potential of Polarized Positrons at the Jefferson Laboratory}

\runningheads{Preparation of Papers for Heron Press Science Series Books}{E. Voutier}

\begin{start}

\author{E. Voutier}{}

\index{Voutier, E.}

\address{Laboratoire de Physique Subatomique et de Cosmologie \\
         Universit\'e Grenoble-Alpes, CNRS/IN2P3 \\
         53 avenue des Martyrs, 38026 Grenoble cedex, France}{}


\begin{Abstract}

Charge symmetry in hadronic reactions, either verified or violated, appears to be in some 
circumstances a mandatory guide for the model-independent understanding of the structure and 
dynamics at play. The recent demonstration of the PEPPo concept for the production of 
polarized positrons opens new physics perspectives at the Jefferson Laboratory. Polarized 
positron beams, in complement to existing polarized electron beams, are shown to bring 
multi-Physics opportunities. 

\end{Abstract}


\end{start}


\section{Introduction}

Quantum Electrodynamics (QED) is one of the most powerful quantum physics theories. The highly 
accurate predictive power of this theory allows not only to investigate numerous physics 
phenomenons at the macroscopic, atomic, nuclear, and partonic scales, but also to test the 
validity of the Standard Model. Therefore, QED promotes electrons and positrons as unique 
physics probes, as demonstrated worldwide over decades at different laboratories.

Both from the projectile and target point of views, spin appears nowadays as the finest tool 
for the study of the intimate structure of matter. Recent examples from the Nuclear Physics 
program developed at the Jefferson Laboratory (JLab) include: the measurement of 
polarization observables in elastic electron scattering off the nucleon~\cite{{Jon00},{Gay02},{Puc10}},  that established the unexpected magnitude and behaviour of the proton electric form factor at high momentum transfer; the experimental evidence, in the production of real photons from a polarized 
electron beam interacting with unpolarized protons, of a strong sensitivity to the orientation of the longitudinal polarization of the electron beam~\cite{Ste01}, that opened the investigation of the partonic structure of the nucleon via the generalized parton distributions~\cite{Mul94} measured through the deeply virtual Compton scattering~\cite{{Ji97},{Rad97}}; etc. Undoubtably, polarization became an important capability and a mandatory property of the current and next accelerator generation.

Combination of the QED predictive power and the fineness of the spin probe led to a large but 
yet limited variety of impressive physics results. Adding to this tool-kit charge symmetry properties 
in terms of polarized positron beams can potentially provide a more complete and accurate picture of 
the physics at play, independently of the size scale involved. In the context of the experimental study of the structure of hadronic matter worked-out at JLab, the comparison between polarized electron and polarized positron induced reactions takes advantage of the lepton beam charge and polarization to  isolate peculiar features of physics revealed by the interference of different QED based reaction  amplitudes.

This contribution reviews the potential benefits of a polarized positron beam at JLab on the basis 
of selected examples of interest for the study of the nucleon structure as well as the 
investigation of the Physics beyond the Standard Model. Opportunities for a polarized thermal positron facility are also described. A more complete and extensive discussion of these benefits 
can be found in  Ref.~\cite{JPos09}.


\section{Elastic scattering}

The measurement of the electric form factor of the nucleon ($G_{E}$) at high momentum transfer, in 
the perspective of the experimental assessment of perturbative Quantum Chromodynamics (QCD) scaling 
laws~\cite{Bro81}, motivated an intense experimental effort targeted by the advent of high energy continuous polarized electron beams. Indeed, the polarization observables technique~\cite{{Akh74},{Arn81}} is expected to be more sensitive to $G_E$  than the cross section method relying on a Rosenbluth separation~\cite{Ros50}. However, the strong disagreement 
between the results of these two experimental methods came as a real surprise. Following the very first measurements of polarization transfer observables in the $^1$H$({\vec e},e{\vec p})$ reaction~\cite{Jon00}, the validity of the Born approximation for the description of the elastic scattering of electrons off protons was questioned. The eventual importance of higher orders in the $\alpha$-development of the electromagnetic interaction was suggested~\cite{Gui03} as an hypothesis to reconciliate cross section and polarization transfer experimental data, while making impossible a model-independent experimental determination of the nucleon electromagnetic form factors via solely electron scattering.

Considering the possible existence of second order contributions to the electromagnetic current, 
the so-called 2$\gamma$-exchange, the $ep$-interaction is no longer characterized by 2 real form factors 
but by 3 generalized complex form factors
\begin{equation}
\widetilde G_{M} = G_{M} + \lambda \, \delta {\widetilde G}_M \,\,\,\,\,\,\, 
\widetilde G_{E} = G_{E} + \lambda \, \delta {\widetilde G}_E \,\,\,\,\,\,\, 
\widetilde F_{3} = \delta {\widetilde F}_3 
\end{equation}
where $\lambda$ represents the lepton beam charge. These expressions involve up to 8 unknown quantities that should be recovered from experiments~\cite{Rek04}. Considering unpolarized electrons, the non point-like structure of the nucleon is expressed by the reduced cross section 
\begin{equation}
\sigma_R = G^2_{M} + \frac{\epsilon}{\tau} \, G^2_{E} - 2 \lambda \, G_{M} \, \Re e \left[ {\mathcal F}_{0} \right] - 2 \lambda \, \frac{\epsilon}{\tau} \, G_{E} \, \Re e \left[ {\mathcal F}_{1} \right]
\end{equation}
where $\mathcal{F}_0 \equiv f_0 \left( \delta {\widetilde G}_M , \delta {\widetilde F}_3 \right)$ and 
$\mathcal{F}_1 \equiv f_1 \left( \delta {\widetilde G}_E , \delta {\widetilde F}_3 \right)$ are additional contributions originating from the 2$\gamma$-exchange mechanisms. The Rosenbluth method, consisting in 
the measurement of the reduced cross section at different $\epsilon$ while keeping constant the momentum transfer, allows the determination of a combination of 1$\gamma$ and 2$\gamma$ electromagnetic form factors and requires consequently model-dependent input to separate further the electric and magnetic 
form factors. The transfer of the longitudinal polarization of an electron beam via the  elastic scattering off a nucleon provides 2 additional and different linear combinations of the same quantities in the form of the transverse ($P_t$) and longitudinal ($P_l$) polarization components of 
the nucleon 
\begin{eqnarray}
& \sigma_R \, P_{t} = - \, P_b \, \sqrt{\frac{2\epsilon(1-\epsilon)}{\tau}} 
\left( G_{E} G_{M} - \lambda \, G_E \Re e \left[ \delta {\widetilde G}_M \right] - \lambda \, 
G_{M} \Re e { \left[ \mathcal{F}_0 \right]} \right) \\
& \sigma_R P_{l} = P_b \, \sqrt{1-\epsilon^2} \left( G_{M}^2 - 2 \lambda \, 
G_{M} \Re e \left[  \mathcal{F}_2 \right] \right) \, .
\end{eqnarray}
where $P_b$ is the lepton beam polarization and $\mathcal{F}_2 \equiv f_2 \left( \delta {\widetilde G}_M , \delta {\widetilde F}_3 \right)$. The combination of single polarized and unpolarized observables of elastic electron scattering involves up to 6 unknown quantities, requiring 6 independent experimental observables.

Once 2$\gamma$-exchange mechanisms are taken into account, polarized electron beams alone cannot provide 
a fully experimental determination of  the electromagnetic form factors of the nucleon. The separation of the 1$\gamma$ and 2$\gamma$ form factors can only be achieved via the sensitivity of experimental  observables to the lepton beam charge. The measurement of polarized and unpolarized elastic scattering of both electrons and positrons would provide the necessary data to allow for the unambiguous separation of the different contributions to the electromagnetic current. Therefore, a model independent determination of the nucleon electromagnetic form factors demands availability of both polarized electron and positron beams.


\section{Deep inelastic scattering}

The understanding of the partonic structure and dynamics of hadronic matter remains a major goal of modern Nuclear Physics. The availability of high intensity continuous polarized electron beams with high 
energy together with performant detector systems at different facilities is providing today an  unprecedented but still limited insight into this problem. Similarly to the elastic scattering case, the combination of measurements with polarized electrons and polarized positrons in the deep inelastic regime 
would allow to obtain model-independent answers to some specific questions.

The generalized parton distributions (GPD) framework~\cite{Mul94} constitutes the most appealing and advanced parameterization of the hadron structure. It encodes the intimate structure of matter in terms of quarks and gluons and unifies within the same framework electromagnetic form factors, parton distributions, and the description of the nucleon spin (see \cite{Die03} for a review). GPDs can be interpreted as the probability to find at a given transverse position a parton carrying a certain 
fraction of the longitudinal momentum of the nucleon. The combination of longitudinal and transverse degrees of freedom is responsible for the richness of this universal framework.
 
GPDs are involved in any deep process and are preferentially accessed in hard lepto-production of real 
photons i.e. the deeply virtual Compton scattering (DVCS). This process competes with the known  Bethe-Heitler (BH) reaction~\cite{Bet34} (recently revisited in Ref.~\cite{Bar13}) where real photons 
are emitted from initial or final leptons instead of the  probed hadronic state. The lepton beam charge and polarization dependence of the cross section off nucleons writes~\cite{Die09}
\begin{equation}
\sigma_{{P_b}0}^{\lambda} = \sigma_{BH} + \sigma_{DVCS} + P_b \, 
\widetilde{\sigma}_{DVCS} + \lambda \, ( \sigma_{INT} + P_b \, \widetilde{\sigma}_{INT} )
\end{equation}
where the index $INT$ denotes the interference between the BH and DVCS amplitudes. Polarized electron 
scattering provides the reaction amplitude combinations
\begin{eqnarray}
\sigma_{00}^{-} & = & \sigma_{BH} + \sigma_{DVCS} - \sigma_{INT} \\
\sigma_{+0}^{-} - \sigma_{-0}^{-} & = & 2 \, \widetilde{\sigma}_{DVCS} - 2 \, \widetilde{\sigma}_{INT} 
\end{eqnarray}
involving four different combinations of the nucleon GPDs. The comparison between polarized electron 
and polarized positron reactions provides the additional combinations
\begin{eqnarray}
\sigma_{00}^{+} - \sigma_{00}^{-} & = & 2 \, \sigma_{INT} \\
\left[ \sigma_{+0}^{+} - \sigma_{+0}^{-} \right] - \left[ \sigma_{-0}^{+} - \sigma_{-0}^{-} \right] 
& = & 4 \, \widetilde{\sigma}_{INT} \, .
\end{eqnarray}
Consequently, measuring real photon lepto-production off the nucleon with opposite charge polarized leptons allows to isolate the 4 unknown contributions to the $ep\gamma$ cross section. 

The separation of nucleon GPDs~\cite{Bel02} requires additional polarization observables measured with polarized targets ($S$). The full lepton beam charge and polarizations dependence of the cross section 
can be generically written~\cite{Die09}  
\begin{eqnarray}
& \sigma_{{P_b} S}^{\lambda} = \sigma_{{P_b} 0}^{\lambda} + \\ & S \left[ P_b \, \Delta\sigma_{BH} + ( \Delta\widetilde{\sigma}_{DVCS} 
+ P_b \, \Delta\sigma_{DVCS} ) + \lambda \, ( \Delta\widetilde{\sigma}_{INT} + P_b \, \Delta\sigma_{INT} ) \right] \nonumber
\end{eqnarray}
where $\Delta\sigma_{BH}$ is the known sensitivity of the BH process to the target polarization and the remaining terms feature four new combinations of the nucleon GPDs to be separated. Polarized electron scattering provides the combinations
\begin{eqnarray}
\sigma_{0+}^{-} - \sigma_{0-}^{-} & = & 2 \, \Delta \widetilde{\sigma}_{DVCS} - 2 \, \Delta \widetilde{\sigma}_{INT}  \\
\left[ \sigma_{++}^{-} - \sigma_{+-}^{-} \right] - \left[ \sigma_{-+}^{-} - \sigma_{--}^{-} \right] & = & 4 \, \Delta\sigma_{BH} + 4 \, \Delta\sigma_{DVCS} - 4 \, \Delta\sigma_{INT} \,\,\,\,\,\,\,\,\,\,\,
\end{eqnarray}
and the comparison between polarized electrons and positrons yields
\begin{eqnarray}
& \left[ \sigma_{0+}^{-} - \sigma_{0-}^{-} \right] - \left[ \sigma_{0+}^{-} - \sigma_{0-}^{-} \right] = 4 \, \Delta \widetilde{\sigma}_{INT}  \\
& \left\{ \left[ \sigma_{++}^{+} - \sigma_{++}^{-} \right] - \left[ \sigma_{-+}^{+} - \sigma_{-+}^{-} \right] \right\}
- \left\{ \left[ \sigma_{+-}^{+} - \sigma_{+-}^{-} \right] - \left[ \sigma_{--}^{+} - \sigma_{--}^{-} \right] \right\} \nonumber \\
& = 8 \, \Delta\sigma_{INT}
\end{eqnarray}
allowing to isolate the four reaction amplitudes of interest. Polarized positron beams then appear as a mandatory complement to polarized electron beams for a model independent determination of nucleon GPDs.


\section{Search for light dark matter boson}

While most of the experimental effort for the search of super symmetry is focussed on heavy 
particle candidates in the TeV mass range, other scenarios involving lighter gauge bosons have also been proposed, consistently with the anomalies observed in cosmic radiations or the anomalous magnetic moment of the muon. The eventual pertinence of a U(1) symmetry, proposed many years ago as a natural extension of the Standard Model~\cite{Fay80}, is suggesting the existence of a new light {\bf U}-boson in the few MeV/$c^2$-GeV$/c^2$ mass range, often referred to as light dark matter~\cite{Fay04} considering its ability to decay through dark matter as well as standard model particles. 

\begin{figure}[th]
\centering
\includegraphics[scale=0.255,angle=0]{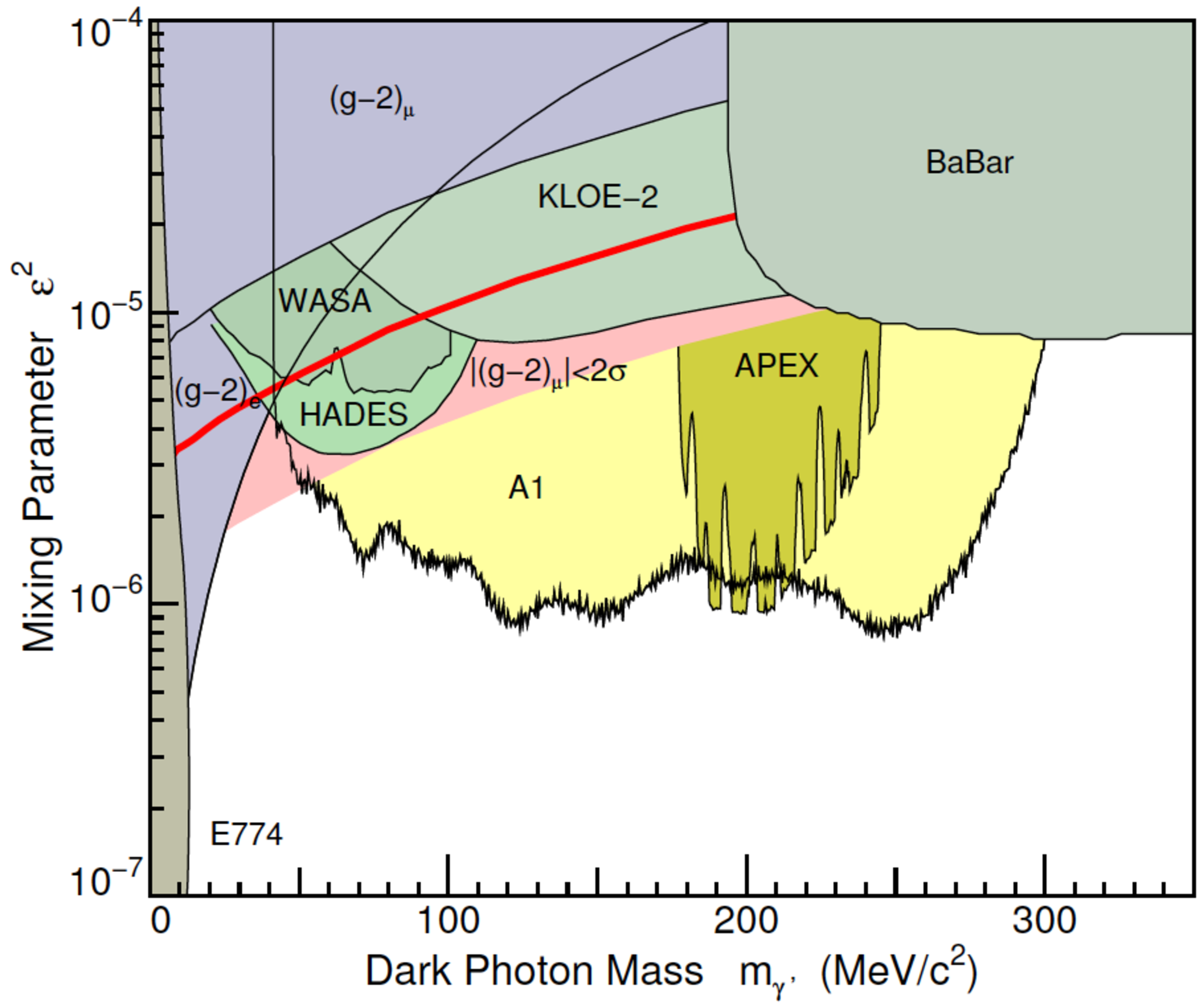}
\caption[]{Experimental status of the exclusion diagram for the search of light dark matter boson~\cite{Mer14}. The unshaded region shows the remaining possible values for the coupling 
strength, relative to the electron charge, as function of the dark photon mass.}\label{dargam}
\end{figure}

The possible existence of such a low mass boson motivated the recent development of several experiments aiming at exploring specific mass and coupling constant ranges in experiments with low energy electron beams: the APEX~\cite{Abr11}, HPS~\cite{Hol11}, and DarkLight~\cite{Fis11} experiments at JLab, as well as dedicated studies at the MAMI~\cite{{Mer14},{Mer11}} and MESA accelerator complex.  Using different techniques, these experiments propose to investigate the reaction $ep \to e p \mathbf{U} \to e p \, e^+ e^-$ featuring the direct production of the {\bf U}-boson and its subsequent decay in an $e^+e^-$ pair. The typical expected signal is a sharp peak at the {\bf U}-boson mass growing  on top of a complex but calculable QED radiative background~\cite{Ber13} and possible hadronic induced background. The most recent status of this difficult experimental search is shown on Fig.~\ref{dargam}, which is  restricting to the low mass region the possible candidates allowing to reconciliate a U(1) extented  Standard Model with $(g$-$2)_\mu$ data. 

Collider like experiments involving reactions between a positron beam and atomic or beam electrons are expected to be much more sensitive and clean probes of the {\bf U}-boson search~\cite{Woj09}. Limiting  the center of mass energy below the pion threshold would guarantee QED based background and would allow investigations of the {\bf U}-boson production through different channels ($e^+ e^- \to \mathbf{U}$, 
$e^+ e^- \to \mathbf{U} \gamma$...) and experimental methods (invisible particle method, invariant mass, missing mass...). Bringing polarization in these experiments can potentially provide additional filtering 
and sensitivity to this search.


\section{Material Science Physics opportunities}

Positron Annihilation Spectroscopy (PAS) is a well-known technique for the investigation of the 
structure of materials~\cite{JPos09} that would benefit from the development of a high intensity 
polarized positron source. PAS is particularly sensitive to the electron density allowing to 
investigate experimentally local structures embedded in the bulk of a material, such as missing atoms, clustering of atoms, superlattices, device structures, quantum dots, voids~\cite{Kra99}... 

\begin{figure}[th]
\centering
\includegraphics[scale=0.235,angle=0]{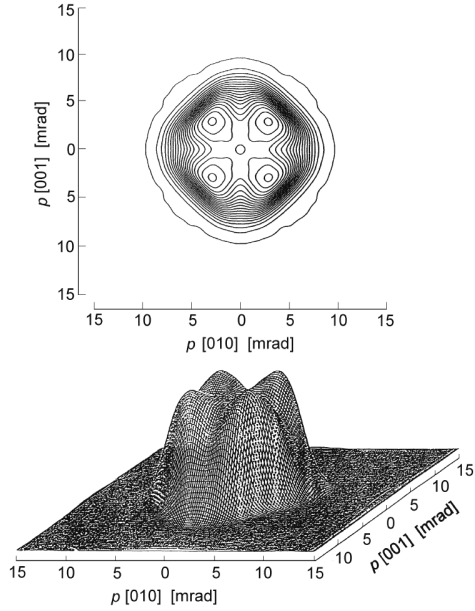}
\caption[]{Two-dimensional angular correlation of annihilation radiation in gallium arsenide 
exhibiting no positron trapping in defects~\cite{Tan95}.}\label{pas2D}
\end{figure}

PAS relies on the annihilation of very low energy positrons with atomic electrons  
and the subsequent detection of one or both of the pair of the annihilation-generated photons. Low
energy positrons penetrating a material, thermalize via multiple scattering and get trapped inside regions with electron density vacanties till they annihilate. The decay time of this process 
is directly related to the electron density at the annihilation site. Furthermore, the motion of 
atomic electrons induce a Doppler broadening of the 511~keV $\gamma$-rays and a distortion of the back-to-back angular correlation. Consequently, the measurement of the energy distribution, or the 
angular correlation between annihilation $\gamma$-rays allows to characterize the material via the determination of the momentum distribution of atomic electrons (Fig.~\ref{pas2D}). However, this powerful technique, known as 2D-ACAR, is limited by the intensity of the available positron beams, which are typically obtained from radioactive sources. The generation of positrons from low energy polarized electrons would be capable to deliver a positron flux about 100 times higher~\cite{Ang14}.  Together with polarization capabilities, such an accelerator based thermalized positron source would be 
a technological breakthrough for PAS studies.


\section{Polarized positrons production}

The production of polarized positrons at the next accelerator generation relies on the creation of $e^+e^-$ pairs by circularly polarized photons produced either from the Compton backscattering of a polarized laser light on a few-GeV electron beam~\cite{Omo06}, or the synchrotron radiation of a multi-GeV electron beam travelling through a helical undulator~\cite{Ale08}. A new approach~\cite{PEPPo} is developed at JLab based on the bremsstrahlung production of circularly polarized photons from polarized electrons. As opposed to other schemes, the PEPPo (Polarized Electrons for Polarized Positrons) concept is potentially capable of operating at lower beam energies (a few-MeV$/c$) with high polarization transfer from incident electrons to created positrons~\cite{Kur10}.

\begin{figure}[th]
\centering
\includegraphics[scale=0.305,angle=0]{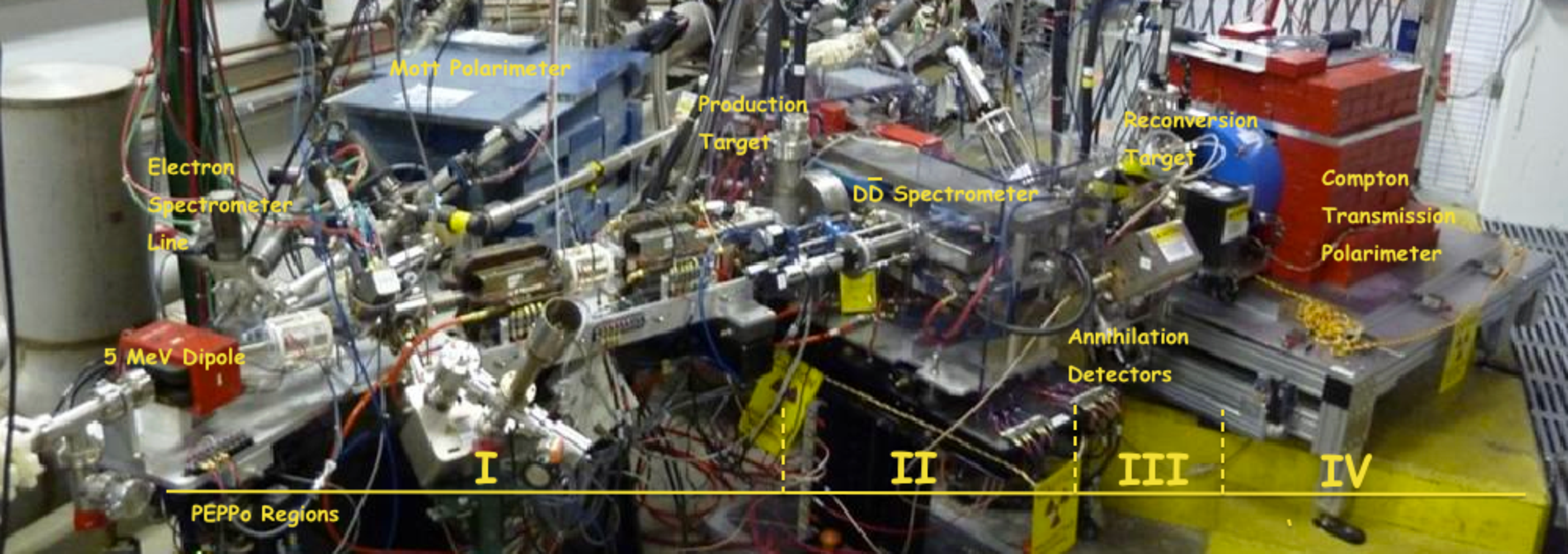}
\caption[]{The PEPPo experimental set-up decomposed in its different regions described in the text.}\label{peppo}
\end{figure}

The PEPPo experiment~\cite{PEPPo} at the injector of the Continuous Electron Beam Accelerator Facility (CEBAF) of JLab has been designed to evaluate and demonstrate the PEPPo concept for a polarized positron source. A few-$\mu$A highly polarized (85\%) continuous electron beam of 8.25~MeV/$c$ is transported to a tungsten production target (0.1-1.0~mm) where polarized positrons are produced (Region I of Fig.~\ref{peppo}). The collection and momentum selection of these positrons 
is insured by a strong solenoid lens followed by $\pm$90$^\circ$ bend $\mathrm{D} \overline{\mathrm{D}}$ 
spectrometer dipoles (Region II of Fig.~\ref{peppo}). A second solenoid at the exit of the spectrometer 
is optimizing the positron transport (Region III of Fig.~\ref{peppo}) to a Compton transmission polarimeter (Region IV of Fig.~\ref{peppo}). The positron polarization is inferred from the measurement 
of the absorption of polarized photons generated in a 2~mm tungsten radiator and absorbed within a 7.5~cm long polarized iron target. The polarimeter has been characterized~\cite{Ade13} with the well-known CEBAF electron beam, and the positron momentum dependence of the experimental asymmetry has been 
measured. The PEPPo Collaboration reported~\cite{{Vou13},{Gra13}} significant on-line experimental  asymmetries supporting high positron polarizations ($>50$\%).

PEPPo clearly opens the ability to produce polarized positrons with any polarized electron beam driver, 
as long as the beam kinetic energy exceeds the pair production threshold. At low energies, suitable beam currents for Physics would be obtained from the mutual optimization of (among others) the initial electron beam intensity and the production target thickness~\cite{Dum11}. The optimum capture of the produced positrons~\cite{Gol14} for further acceleration (Nuclear and Hadronic Physics) or deceleration (Material Science Physics) is an additional issue which may require supplementary systems like cooling- and/or 
damping-rings and/or cavities. In the context of the JLab accelerator facility, several scenarios can be considered depending on the initial electron beam energy and technological constraints but conceptual studies are already promising: the full lattice simulation of a design based on a 100~MeV$/c$ and 1~mA electron beam driver showed that positron beam intensities up to 300~nA can be expected with a momentum spread of 10$^{-2}$ and a beam emittance $(\varepsilon_x \varepsilon_y) = (1.6,1.7)$~mm.mrad~\cite{Gol10}, that are well within specifications required to perform the proposed physics program~\cite{JPos09}.  


\section{Conclusion}

The combination of polarized electron and positron beams at JLab can potentially resolve in 
a model-independent way fundamental issues of the structure and dynamics of hadronic matter. 
Polarized positron beams therefore appear as an important, and sometimes mandatory, complement 
for the accomplishment of the JLab Nuclear Physics program. Their benefits extend to the 
development of the new Physics via accurate test of the Standard Model, and new capabilities 
via applications of intense polarized thermalized positron beams in Material Science Physics. 
Altogether, these opportunities would constitute an important enhancement of the scientific reach 
of the 12~GeV Upgrade. 


\section*{Acknowledgements}

The author thanks the organizers of XXXIIIrd International Workshop on Nuclear Theory for their
invitation and warm hospitality at Rila Mountains. This work is the combination of many contributions over several years. It is a privilege to thank V.~Angelov, J.~Dumas, J.~Grames and the PEPPo Collaboration for long-standing and fruitful collaboration. This work was supported in part by the French 
GDR 3034 Chromodynamique Quantique et Physique des Hadrons.



\end{document}